\documentclass[fleqn,usenatbib]{mnras}

\usepackage{newtxtext,newtxmath}

\usepackage[T1]{fontenc}

\DeclareRobustCommand{\VAN}[3]{#2}
\let\VANthebibliography\thebibliography
\def\thebibliography{\DeclareRobustCommand{\VAN}[3]{##3}\VANthebibliography}

\usepackage{graphicx}	
\usepackage{amsmath}	

\usepackage{gensymb}

\newcommand{\Mpc}{\rm\; Mpc}
\newcommand{\kpc}{\rm\; kpc}

\newcommand{\km}{\rm\; km}

\newcommand{\cm}{\rm\; cm}

\newcommand{\yr}{\rm\; yr}
\newcommand{\Gyr}{\rm\; Gyr}
\newcommand{\Myr}{\rm\; Myr}
\newcommand{\s}{\rm\; s}

\newcommand{\ks}{\rm\; ks}

\newcommand{\MHz}{\rm\; MHz}

\newcommand{\Msun}{\hbox{$\rm\thinspace M_{\odot}$}}

\newcommand{\Msunpyr}{\hbox{$\Msun\yr^{-1}\,$}}

\newcommand{\keV}{\rm\; keV}
\newcommand{\eV}{\rm\; eV}
\newcommand{\erg}{\rm\; erg}

\newcommand{\ergps}{\hbox{$\erg\s^{-1}\,$}}

\newcommand{\kmps}{\hbox{$\km\s^{-1}\,$}}

\newcommand{\kmpspMpc}{\hbox{$\kmps\Mpc^{-1}\,$}}

\newcommand{\Zsun}{\hbox{$\thinspace \mathrm{Z}_{\odot}$}}

\newcommand{\amin}{\rm\; arcmin}

\newcommand{\asec}{\rm\; arcsec}

\newcommand{\psqcm}{\hbox{$\cm^{-2}\,$}}

\title[Gas motions in a bubble's wake]{The splash beneath the largest radio bubble in a cluster core}

\author[H.~R. Russell et al.]{
H.~R. Russell$^{1}$\thanks{E-mail: helen.russell@nottingham.ac.uk}, P.~E.~J. Nulsen$^{2,3}$, A.~C. Fabian$^4$, B.~R. McNamara$^5$, J.~S. Sanders$^6$, and N. Werner$^7$
\\
$^{1}$School of Physics \& Astronomy, University of Nottingham, University Park, Nottingham NG7 2RD, UK\\
$^{2}$Center for Astrophysics $|$ Harvard \& Smithsonian, 60 Garden Street, Cambridge, MA 02138, USA\\
$^{3}$ICRAR, University of Western Australia, 35 Stirling Hwy, Crawley, WA 6009, Australia\\
$^4$ Institute of Astronomy, Madingley Road, Cambridge CB3 0HA, UK\\
$^5$Waterloo Centre for Astrophysics, University of Waterloo, ON N2L 3G1, Canada\\
$^{6}$ Max-Planck-Institut f{\"u}r extraterrestrische Physik, Gie{\ss}enbachstra{\ss}e 1, D-85748, Garching, Germany\\
$^{7}$Department of Theoretical Physics and Astrophysics, Faculty of Science, Masaryk University, Kotlářská 2, Brno, 611 37, Czech Republic
}

\date{Accepted XXX. Received YYY; in original form ZZZ}

\pubyear{\the\year{}}

\begin{document}
\label{firstpage}
\pagerange{\pageref{firstpage}--\pageref{lastpage}}
\maketitle

\begin{abstract}
We present a $100\ks$ XRISM Resolve observation of the Ophiuchus cluster that measures turbulence and bulk motion in the wake of the largest radio bubble on the sky.  We detect a significant velocity shift of $-80\pm20\kmps$ from the cluster centre to the bubble's wake and a clear increase in velocity dispersion from $135\pm10\kmps$ to $210\pm20\kmps$.  The measured bulk velocity in the wake is low and suggests that the bubble's trajectory is inclined with respect to the line of sight.  If we subdivide the bubble's wake, fitting spectra simultaneously with cross-region responses, we find that the velocity shift and dispersion increase are primarily detected in the very centre of the wake.  This is consistent with the expected updraft, or `splash', found beneath buoyantly rising radio bubbles.  In the cluster's cool core, the turbulent kinetic energy is only $1\%$ of the thermal energy radiated over a cooling timescale of $7\Gyr$, and even falls short, by a factor of 5, of the thermal energy radiated over the bubble's rise time.  Whilst turbulent energy generated in the large wake region may provide additional heating, this propagates too slowly to prevent rapid cooling across the core.  The turbulent-dissipation heating rate is a factor of $\sim3$ below the cooling luminosity.  Despite the vast power of the giant radio bubble in the Ophiuchus cluster, the gas motions in the wake are remarkably modest and turbulent-dissipation appears unable to prevent rapid cooling.
\end{abstract}

\begin{keywords}
galaxies: clusters: intracluster medium -- X-rays: galaxies: clusters -- galaxies: clusters: Ophiuchus
\end{keywords}



\section{Introduction}


The impact of active galactic nuclei (AGN) on massive galaxies and clusters first became apparent in high resolution X-ray images more than two decades ago \citep[for reviews see][]{Fabian12,McNamaraNulsen12}.  These systems are surrounded and pervaded by hot, volume-filling plasma at millions of degrees that radiates X-rays.  Powered by accretion onto the central supermassive black hole, AGN launch relativistic jets that carve out large cavities in their surrounding hot atmospheres (\citealt{Boehringer93,Churazov00,McNamara00}).  These cavities are visible as surface brightness depressions in X-ray images and radio observations reveal that they are filled with relativistic plasma.  The jet energy that is required to evacuate the implied volumes is substantial, typically $10^{43-45}\ergps$ \citep[e.g.][]{Birzan04,Rafferty06,Hlavacek-Larrondo12}, and comparable to the energy radiated by the most luminous quasars.  

If the jets efficiently heat their surrounding environment, this mechanism can prevent "overcooling" in massive galaxies and clusters (\citealt{Pedlar90,Baum91,Tucker97}).  Simulations of galaxy formation, that include only gravity and radiative cooling, typically predict the steady accumulation of far too much cold gas and star formation in massive galaxies.  This reaches an extreme in rich galaxy clusters where cooling flows in hot atmospheres would flood the central galaxy with cold gas at rates of $100-1000\Msunpyr$ (e.g. \citealt{Fabian94}).  Inclusion of AGN heating in cosmological simulations has been shown to be able to suppress gas cooling and to truncate the growth of massive galaxies at late cosmic times, in line with observations \citep[e.g.][]{DiMatteo05,Bower06,Croton06,Hopkins06,Nelson18}.

Most of the energy injected by the jet goes into the
internal energy of the inflated radio bubble and the potential energy
of the displaced hot gas (\citealt{Churazov00, Churazov02, Begelman01}).  As the bubble buoyantly rises, this energy must somehow be transferred to the intracluster medium (ICM) and eventually dissipated as heat.  A number of different mechanisms could facilitate the transfer: turbulence generated in the bubble's wake, uplift of gas entrained by the bubble, weak shocks and sound waves, excitation of internal waves, cosmic rays and mixing of hot gas from the radio bubbles (\citealt{Churazov01,FabianPer03,Ensslin03,Ruszkowski04,Dennis05,Mathews06,Ruszkowski08,Zhuravleva14,Reynolds15,Hillel16,Bambic18,Zhang18}).

Jet expansion and bubble inflation are expected to induce corresponding motions in hot atmospheres.  For powerful jetted outbursts, the expanding bubble pushes the underlying ICM to smaller radii in a backflow.  Later on, the buoyant bubble starts to rise and the pressure eases.  The displaced ICM then accelerates outward, driven by the core pressure around it, and fills the region evacuated by the rising bubble.  The collapse of an evacuated region is known as a `splash' and, in this case, it results in the strong updraft behind the rising bubble.  Until recently, limited spectral resolution at X-ray wavelengths has prevented us from directly measuring gas velocities below a few hundred km/s (\citealt{Sanders11,Pinto15}, for a review see \citealt{Simionescu19}).  Indirect measurements, from studies of surface brightness fluctuations (\citealt{Zhuravleva14,Walker15,Werner16}), resonant scattering (\citealt{Werner09,dePlaa12,Ogorzalek17}) and cooler gas phases (\citealt{McDonald12,Werner14,Hamer16,Gaspari18,Olivares19}), indicated modest gas velocities and turbulent motions.  However, the arrival of X-ray microcalorimeters, onboard Hitomi (\citealt{Hitomi16}) and now XRISM (\citealt{XRISM25Centaurus,XRISM25Perseus,XRISM25M87,Rose25}), finally allows us to directly measure and map the gas dynamics in hot atmospheres.  

Although XRISM's spectral resolution is exceptional, the modest spatial resolution ensures that the best targets for studies of AGN feedback are the nearest, brightest galaxy clusters with active nuclei.  At 13 arcmin across ($460\kpc$), the X-ray cavity in the nearby Ophiuchus cluster is, by far, the largest on the sky \citep[][]{Giacintucci20,Giacintucci25}.  The Ophiuchus cluster is second only to the Perseus cluster in X-ray brightness ($2-10\keV$, \citealt{Edge92}) and is particularly hot ($\sim9\keV$, \citealt{Johnston81,Arnaud87}), which ensures strong Fe K line
emission.  Although there are signatures of gas sloshing within the
cool core, the bubble's wake is clear of projected structure.
Fig. \ref{fig:chandraimg} (left) shows the \textit{Chandra} X-ray image covering the cluster's cool core and, at a radius of 3.5 arcmin ($120\kpc$), a concave, sharp surface brightness edge surrounding a vast cavity
\citep[][]{Werner16}.  Subsequent low frequency radio observations
with the GMRT (Fig. \ref{fig:chandraimg} right) and from MWA/GLEAM revealed that this cavity is filled
with diffuse radio emission and has a very steep radio spectrum
\citep[][]{Giacintucci20,Giacintucci25}.  This structure is therefore an old radio
bubble generated by the most powerful AGN outburst known in a galaxy
cluster.  A counterpart bubble is not apparent in the X-ray or radio
observations but this is likely due to low X-ray surface
brightness and asymmetry at such large radii and radio spectral aging.  Measurements of the gas motions in Ophiuchus, from XMM-Newton spectroscopy (\citealt{Gatuzz23}) and \textit{Chandra} observations of surface brightness fluctuations (\citealt{Werner16}), indicate modest bulk velocities and velocity dispersions.

Here we present a XRISM Cycle 1 observation covering the wake of this vast radio bubble to measure the induced gas motions.  Section \ref{sec:obs} covers the observational setup, data calibration and generation of spectra and responses.  Section \ref{sec:spec} details the spectral modelling, spatial-spectral mixing analysis and resulting measurements of the velocity structure.  Section \ref{sec:disc} discusses the gas motions around the large radio bubble and the energy transfer to the surrounding ICM.

From a dynamical analysis of 142 cluster member galaxies, \citet{Durret15} measure a mean redshift $z_{\mathrm{gal}}=0.0296\pm0.0003$ for the Ophiuchus cluster (\citealt{Durret15}).  The brightest cluster galaxy (BCG) is spatially coincident with the X-ray emission peak and its velocity is consistent with the mean velocity of the cluster members, $\Delta v = -47 \pm 97\kmps$ (\citealt{Durret15}).  We assume $H_0=70\kmpspMpc$, $\Omega_{\mathrm{m}}=0.3$ and
$\Omega_\Lambda=0.7$, translating to a scale of $0.59\kpc$ per arcsec.  All errors are $1\sigma$ unless otherwise noted.

\section{Observations and data reduction}
\label{sec:obs}

\begin{figure*}
\begin{minipage}{\textwidth}
\centering
\includegraphics[width=0.45\columnwidth]{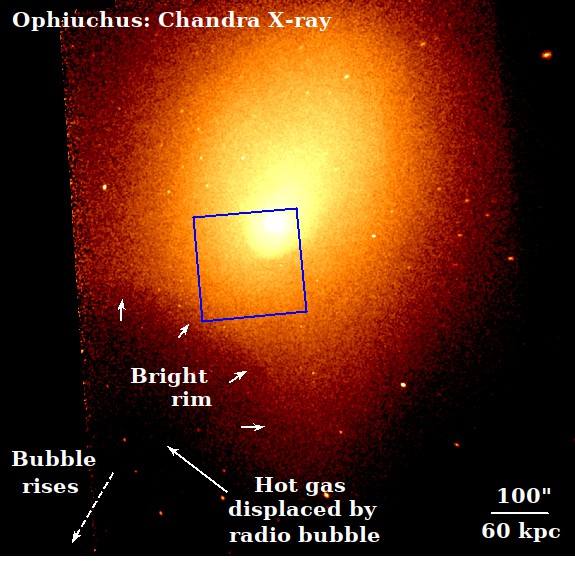}
\raisebox{0.05cm}{\includegraphics[width=0.45\columnwidth]{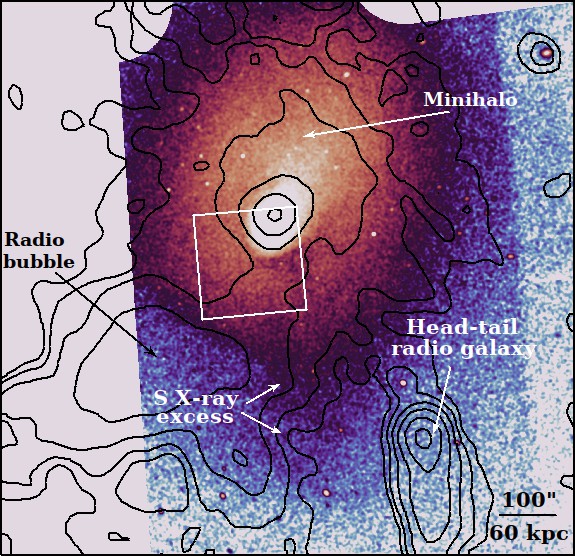}}
\caption{Left: \textit{Chandra} X-ray image ($0.5-7\keV$) of the Ophiuchus cluster showing the X-ray cavity and rim to the SE of the cool core.  The XRISM field of view is shown by the blue region.  Right: With GMRT $210\MHz$ radio contours from \citet{Giacintucci25}.  The X-ray cavity is filled with radio emission.  A radio minihalo is detected in the cluster core and a head-tail radio galaxy appears to the SW.}
\label{fig:chandraimg}
\end{minipage}
\end{figure*}

\begin{figure}
    \centering
    \includegraphics[width=0.9\linewidth]{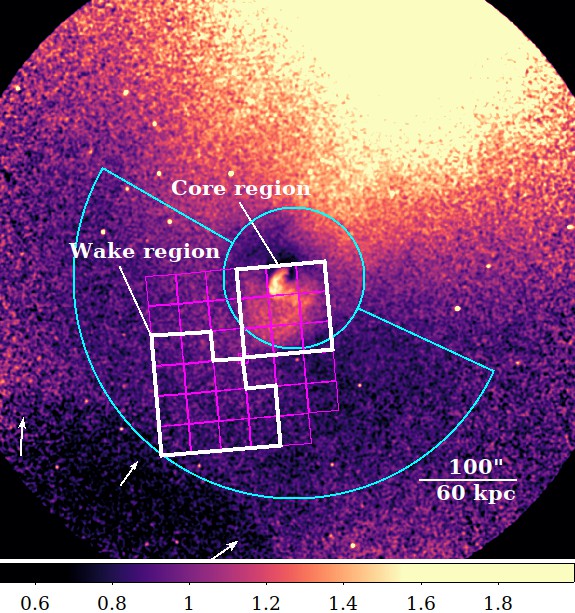}
    \caption{\textit{Chandra} residual image where the surface brightness has been divided by the average value at each radius, which was evaluated in circular annuli that were each $2.5\asec$ in width.  The core and wake detector regions used are shown in white and the sky regions used for the spatial-spectral mixing analysis are shown in cyan.}
    \label{fig:skyreg}
\end{figure}

XRISM observed the bubble wake in the Ophiuchus cluster in a single pointing from March 28 to March 31, 2025 (obs. ID 201117010).  The observation was taken with the aimpoint at $\alpha=258.129\degree$, $\delta=-23.3929\degree$ and with position angle $94.8\degree$.  Fig. \ref{fig:chandraimg} (left) shows the positioning of Resolve's $3.1\arcmin\times3.1\arcmin$ field of view beneath the X-ray cavity detected by \textit{Chandra}.  The $6 \times 6$ pixel array is shown, where each pixel produces a spectrum with a resolution of $4.5\eV$ (FWHM, \citealt{Porter24}).  The corner calibration pixel 12 was excluded from our analysis and is not shown.

The Resolve dataset was analysed with the latest version of HEASoft 6.35.1 and version 11 of the calibration database (CALDB).  The dataset was reprocessed with the pipeline task, \textsc{xapipeline} and standard screening was applied as detailed in the XRISM Data Reduction Guide\footnote{\url{https://heasarc.gsfc.nasa.gov/docs/xrism/analysis/abc_guide/xrism_abc.html}} and in \citet{XRISM25A2029}.  Pixel 27 was excluded because it exhibits unexplained gain excursions that cannot be corrected with the current calibration procedure (\citealt{Porter24}).  The final cleaned exposure time was $116.6\ks$.  

We apply a barycentric velocity correction of $28\kmps$ (e.g. \citealt{Wright14}) to the measured velocities to enable comparison with the optical measurements of cluster galaxy velocities.  The XRISM observation was completed over several orbits therefore the additional orbital motion of the spacecraft will average out and the resulting broadening is not significant.

Spectra were extracted using only the high-primary events, which represent more than 99\% of the $1.8-10.0\keV$ events in this observation and ensure the highest spectral resolution.  Corresponding response files were generated to correct each spectrum for instrumental response.  Large redistribution matrix files (RMFs) were generated with the \textsc{rslmkrmf} task.  Ancillary response files, correcting for the energy-dependent effective area, were generated with \textsc{xaarfgen}.  This task was run in `image' mode with the input brightness distribution provided by a $1.8-8\keV$ \textit{Chandra} image of the Ophiuchus cluster covering the XRISM field of view.  

Due to XRISM's $1.3\amin$ point spread function (\citealt{Tashiro25}), spectra extracted from subregions of the Resolve field of view will contain photons that originate from outside the selected region.  Although spectra extracted for the full Resolve field of view will feature photons from neighbouring regions on the sky, this effect is minor, particularly away from the bright cool core.  For our analysis of detector subregions of the field (white regions in Fig. \ref{fig:skyreg}), we determine the contributed emission from each of two sky regions (blue regions in Fig. \ref{fig:skyreg}) in a spatial-spectral mixing analysis.  The region sizes were determined by XRISM's spatial resolution and a requirement for enough photons to constrain the velocity broadening.  The sky regions cover the bright cool core and the wake region and neighbouring regions at comparable brightness that contribute scattered photons.  We used a circular region of radius $1.2\amin$ to cover the bright cool core and a sector from $1.2$ to $3.75\amin$ to cover the wake region.  The spectra from each detector region were fitted simultaneously with appropriate cross-region responses (\citealt{Hitomi18}).


A non-X-ray background (NXB) spectrum was extracted from the Resolve NXB database of `night Earth' data using the \textsc{rslnxbgen} task.  The spectral flux of the NXB was more than an order of magnitude fainter than the cluster emission over the full energy band $1.8-10.0\keV$ analysed.  The impact on our spectral fit results for the full field of view was found to be negligible.  The cosmic X-ray background, from unresolved distant AGN, is even fainter than the NXB.  Therefore, following e.g. \citet{XRISM25A2029}, we ignored the contribution of the total background to our spectral fits.

\section{Spectral analysis}
\label{sec:spec}

\begin{figure*}
\begin{minipage}{\textwidth}
\centering
\includegraphics[width=0.95\columnwidth]{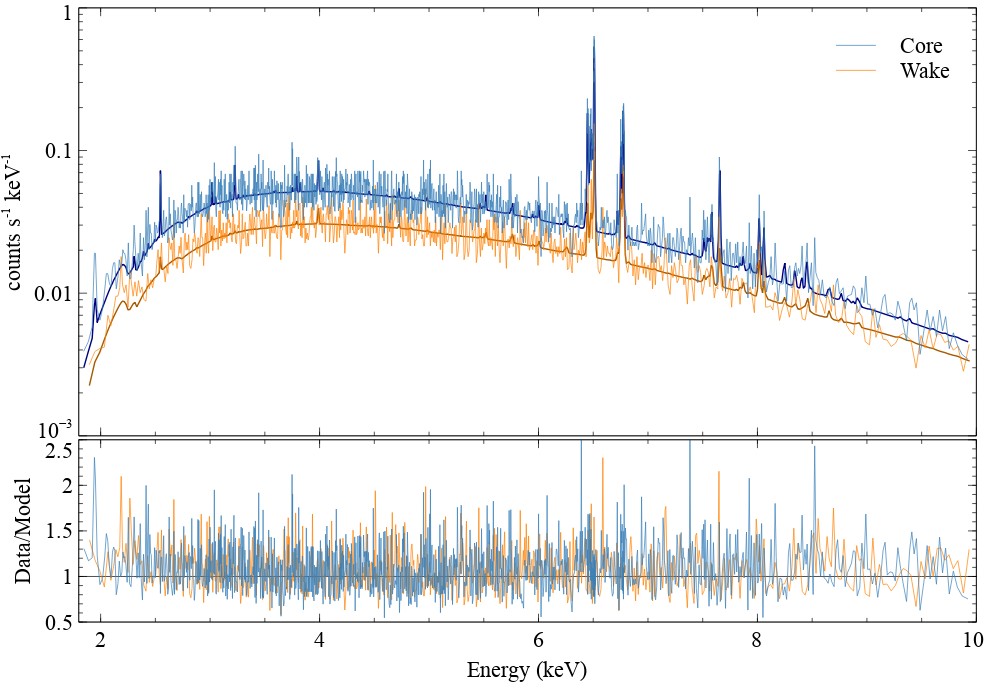}
\caption{Resolve spectra, models and ratios for the core and wake regions shown in Fig. \ref{fig:skyreg}.  The spectra are shown grouped to a minimum of 20 counts per bin for clarity but all spectral fitting was performed with a minimum of 1 count per bin.}
\label{fig:fullspec}
\end{minipage}
\end{figure*}

Spectra were grouped with a minimum of one count per bin and modelled in \textsc{xspec} version 12.15.0 (\citealt{Arnaud96}) with atomic line emission models from AtomDB version 3.0.9.  The model was fitted over a broad energy band, $1.8-10.0\keV$ with an absorbed single temperature thermal plasma emission model \textsc{tbabs(bvapec)}.  The absorption is due to gas in the Milky Way along the line of sight to the Ophiuchus cluster.  Although Ophiuchus is located at a low Galactic latitude, and therefore strongly absorbed at soft energies, the impact is minimal above $2\keV$.  We therefore fixed the absorbing column to the Galactic value of $1.9\times10^{21}\psqcm$ (\citealt{HI4PI16}).  Abundances in the cluster gas were measured relative to the proto-solar abundances of \citet{Lodders09}.  The abundances of He and C were fixed to Solar.  The S, Ar, Ca, Fe and Ni abundances were left free and the rest were tied to Fe.  The temperature, redshift, velocity broadening and normalization parameters were left free.  \textsc{xspec}'s version of the C-statistic was minimised (\citealt{Cash79,Wachter79}).  


Single temperature thermal plasma models provided good fits to the data in all regions, once spatial-spectral mixing was accounted for.  We found no evidence of resonant scattering in the bright core (or the wake), in agreement with \citet{Fujita25}.  We also tested the best-fit models for a narrow energy band covering only the brightest emission lines from $6$ to $7\keV$.  We froze the abundances for elements with emission lines located outside this range.  The narrow band results were found to be entirely consistent with the broad band results for both the core and wake regions.  We therefore proceeded with the broad band results.


\subsection{Core and wake regions}
\label{sec:coreandwake}

\begin{figure*}
\begin{minipage}{\textwidth}
\centering
\includegraphics[width=0.45\columnwidth]{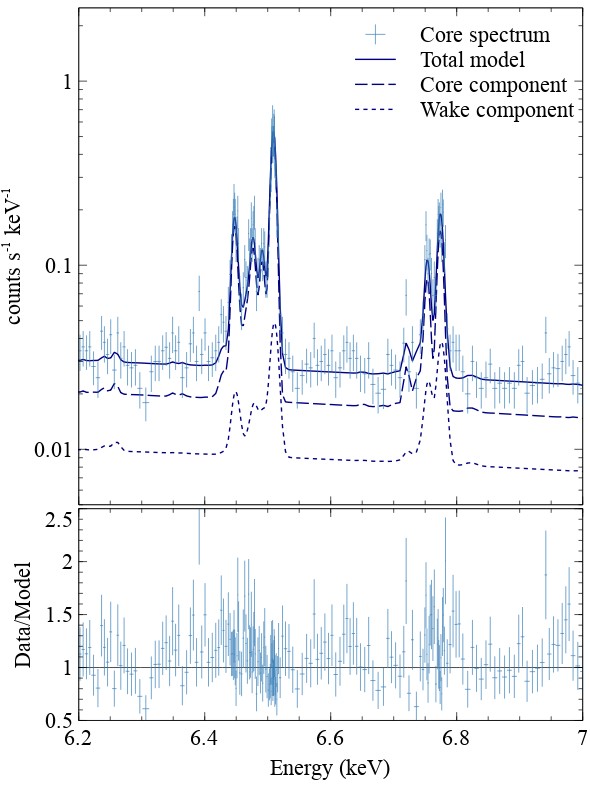}
\includegraphics[width=0.45\columnwidth]{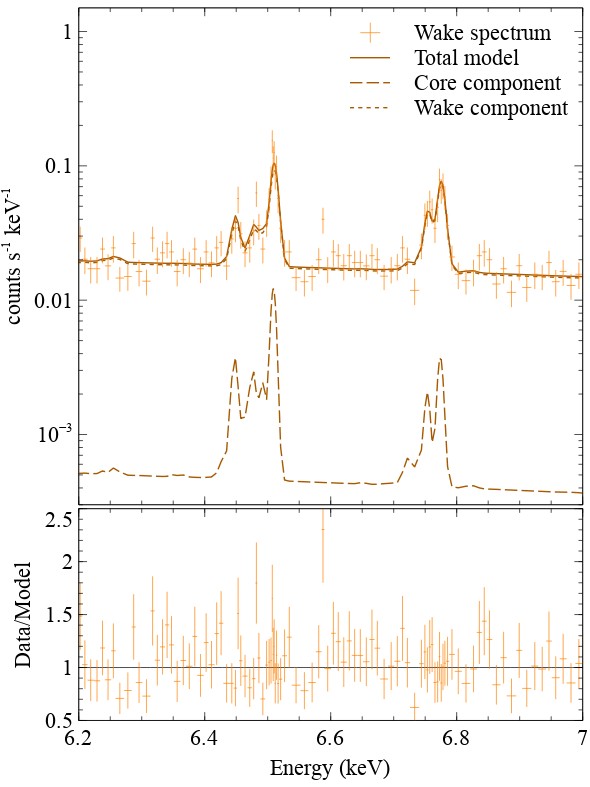}
\caption{Resolve spectra, models and ratios for the core and wake regions shown in Fig. \ref{fig:skyreg}.  The spectra are zoomed to show the Fe XXV He$\alpha$ and Fe XXVI Ly$\alpha$ emission lines and the contributions of the spatial-spectral mixing components determined using each sky region are shown.  The spectra are shown grouped to a minimum of 20 counts per bin for clarity but all spectral fitting was performed with a minimum of 1 count per bin.}
\label{fig:fullspec_split}
\end{minipage}
\end{figure*}

Fig. \ref{fig:fullspec} shows the Resolve spectra extracted from the core and wake regions (Fig. \ref{fig:skyreg}).  The spectra were fitted simultaneously in \textsc{xspec} using the cross-region responses as discussed in section \ref{sec:obs}.  The relative contributions of each spectral component are shown in Fig. \ref{fig:fullspec_split}.  Scattered photons from the core make only a modest contribution in the large wake region.  However, the smaller core region does contain a significant fraction of photons from the surrounding sky region that includes the bubble's wake.

The different shapes of the Fe XXV He$\alpha$ and Fe XXVI Ly$\alpha$ emission lines in Fig. \ref{fig:fullspec_split} highlight the differences in the gas properties between these spatial regions.  The core is significantly cooler, at $6.3\pm0.1\keV$, than the surrounding gas at $9.4\pm0.3\keV$.  The core is also metal-rich with an Fe abundance of $0.73\pm0.03\Zsun$ due to stellar activity in the central massive galaxy.  This is significantly enhanced over the surrounding gas with an Fe abundance of $0.33^{+0.03}_{-0.02}\Zsun$.  

The bulk velocity along the line of sight is given by

\begin{equation}
v_{\mathrm{los}} = \frac{c(z_{\mathrm{ICM}} - z_{\mathrm{gal}})}{1+z_{\mathrm{gal}}}
\end{equation}

\noindent where $z_{\mathrm{ICM}}$ is the best-fit redshift for the cluster's atmosphere.  For the core region, the bulk velocity is $-36\pm5\kmps$ and the stated uncertainty is the statistical uncertainty from Resolve spectral fitting. The bulk velocity of the gas is therefore modestly offset from the mean velocity of the cluster members (although consistent with the central galaxy's velocity within the associated larger uncertainties).  The velocity dispersion in the core is low at $135\pm10\kmps$.  Our results for the core region are essentially consistent with the findings of \citet{Fujita25}, who analysed a XRISM observation centred on the Ophiuchus cluster's emission peak.  Our core region is slightly offset from X-ray peak, which accounts for modest discrepancies in the core temperature and gas motions.  All best-fit parameters are given in Table \ref{tab:specpars}.


\begin{table}
\caption{The best-fit spectral parameters for the core and wake regions in the Ophiuchus cluster.}
\begin{center}
\begin{tabular}{ccc}
\hline
Parameter & Core & Wake \\
\hline
Temperature (keV) & $6.3\pm0.1$ & $9.4\pm0.3$ \\
Fe abundance (\Zsun) & $0.73\pm0.03$ & $0.33^{+0.03}_{-0.02}$\\
S abundance (\Zsun) & $1.0\pm0.2$ & $0.4\pm0.3$\\
Ar abundance (\Zsun) & $1.1\pm0.3$ & $0.1^{+0.4}_{-0.1}$\\
Ca abundance (\Zsun) & $0.9\pm0.3$& $0.4\pm0.4$\\
Ni abundance (\Zsun) & $0.8\pm0.2$ & $0.3\pm0.2$\\
Velocity, $v_{\mathrm{los}}$ (km/s) & $-36\pm5$ & $-120\pm20$\\
Dispersion, $\sigma$ (km/s) & $135\pm10$ & $210\pm20$\\
C-stat/dof & 10374/11852 & 7685/9672\\
\hline
\end{tabular}
\label{tab:specpars}
\end{center}
\end{table}

\begin{figure}
\centering
\includegraphics[width=0.85\columnwidth]{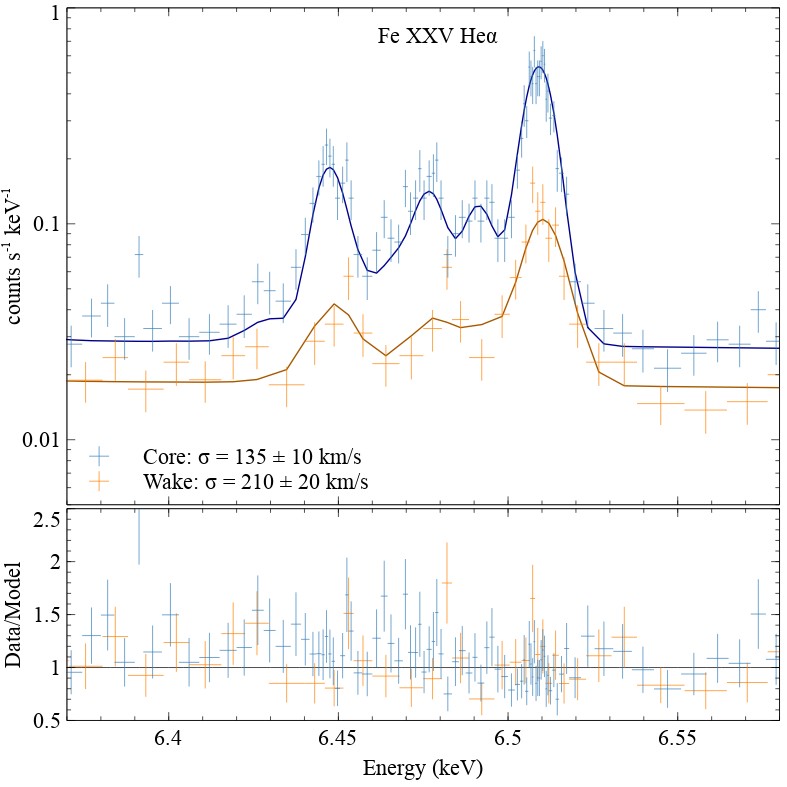}
\includegraphics[width=0.85\columnwidth]{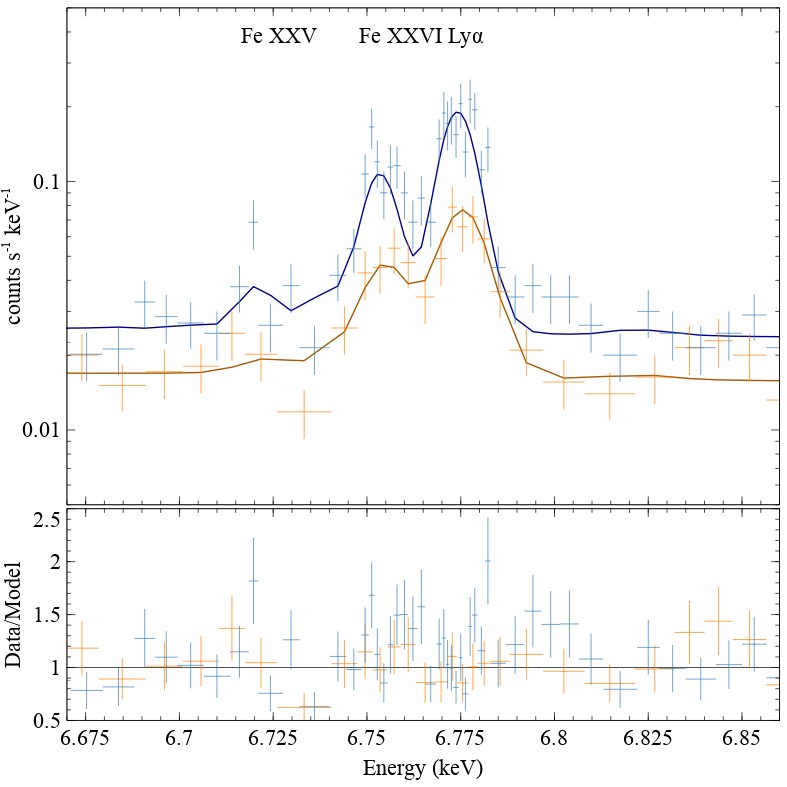}
\includegraphics[width=0.85\columnwidth]{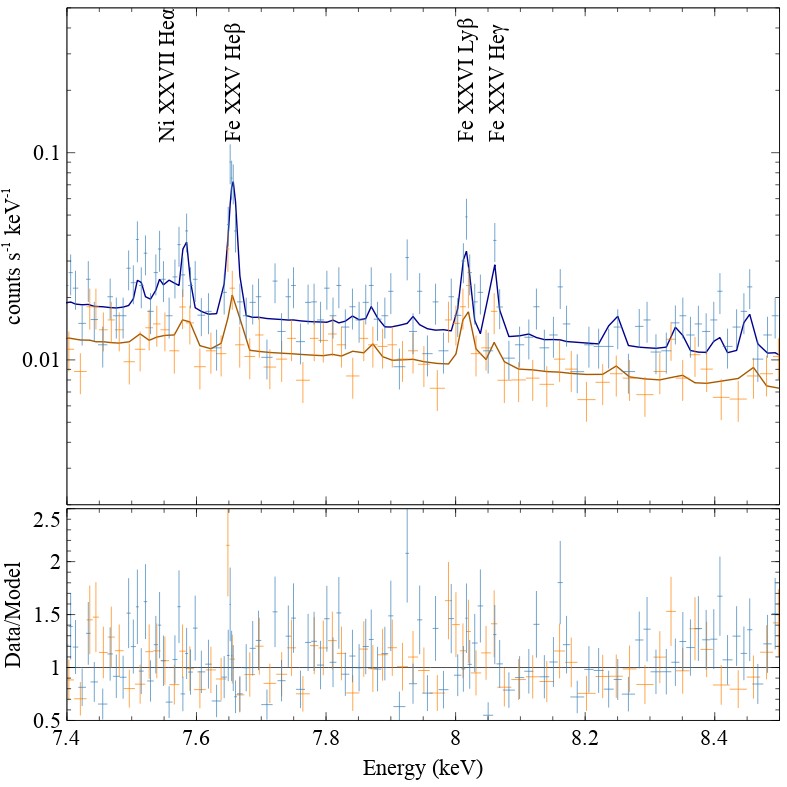}
\caption{Resolve spectra, models and ratios for the core and wake regions shown in Fig. \ref{fig:skyreg}.  The spectra are zoomed to the brightest line complexes and shown grouped to a minimum of 20 counts per bin for clarity.}
\label{fig:subspec}
\end{figure}

We detect a significant velocity shift and increase in velocity dispersion in the wake region.  Fig. \ref{fig:subspec} shows the broader Fe XXV He$\alpha$ and Fe XXVI Ly$\alpha$ emission lines in the wake compared to the core and a modest shift in the line centroids to higher energies.  We measure a bulk velocity in the wake of $-120\pm20\kmps$ and a dispersion of $210\pm20\kmps$.  The gas temperature increases sharply by $\sim50\%$ in this region to $9.4\pm0.3\keV$ and the Fe abundance drops to $0.33^{+0.03}_{-0.02}\Zsun$.  \textit{Chandra} observations reveal a sharp cold front between the core and wake regions with an isothermal hot medium beyond the truncated cool core (\citealt{Werner16}).  XRISM now reveals that the gas motions are also distinct with a significant shift in bulk velocity and increase in velocity dispersion.

\subsection{Subdividing the bubble's wake}

\begin{figure}
\centering
\includegraphics[width=0.9\columnwidth]{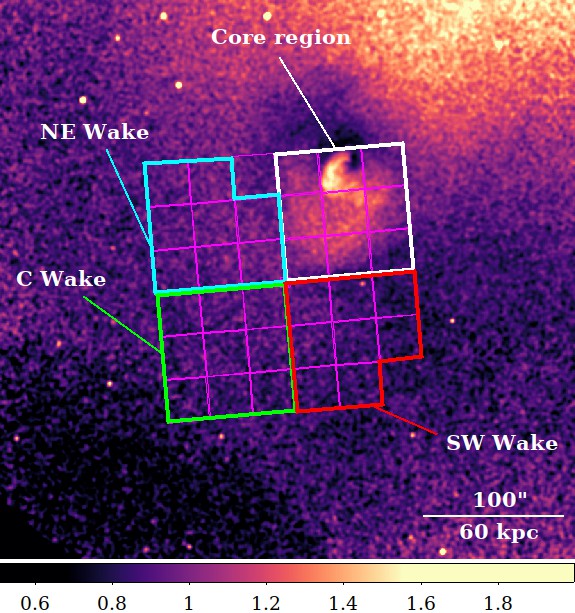}
\caption{\textit{Chandra} residual image (as in Fig. \ref{fig:skyreg}) with the core and subdivided wake regions shown.}
\label{fig:chandrasubdiv}
\end{figure}

We repeated our analysis with spectra extracted from the core and three separate regions spanning the bubble's wake (Fig. \ref{fig:chandrasubdiv}).  All spectra were fitted simultaneously in \textsc{xspec} with appropriate cross-region responses.  Based on the remarkably isothermal and uniform metal distribution outside the cool core (\citealt{Werner16}), we tied the temperature and metallicity parameters across the three wake regions.  The redshift and velocity dispersion parameters for the four \textsc{bvapec} components were left free.  The increased uncertainties reflect the addition of two model components and the significant fractions of scattered light between the three wake regions.  The best-fit parameters are detailed in Table \ref{tab:specdivpars}.  The best-fit temperature and metallicity values determined in the core and tied across the wake are consistent with the best-fit values for the core and wake regions determined in Section \ref{sec:coreandwake} and detailed in Table \ref{tab:specpars}.  The velocity shift and increase in velocity dispersion in the wake are primarily observed in the wake centre region.  The velocity shift is also detected in the SW region but the NE region is consistent with the core motion.


\begin{table}
\caption{The best-fit spectral parameters for the core and subdivided wake regions in the Ophiuchus cluster.}
\begin{center}
\begin{tabular}{cccc}
\hline
Region & Velocity $v_{\mathrm{los}}$ & Dispersion $\sigma$ & C-stat/dof \\
 & (km/s) & (km/s) & \\
\hline
Core & $-37\pm5$& $140\pm10$ & 10378/11852 \\
Wake NE & $-40^{+20}_{-10}$& $170\pm30$ & 7006/8698 \\
Wake C & $-180^{+50}_{-60}$ & $240\pm60$ & 5850/7604 \\
Wake SW & $-180\pm30$ & $140\pm40$ & 6496/8164 \\
\hline
\end{tabular}
\label{tab:specdivpars}
\end{center}
\end{table}


\section{Discussion}
\label{sec:disc}

\subsection{Terminal velocity of the radio bubble}
\label{sec:terminal}

Simulations of buoyant bubbles in the ICM typically show that hot gas is pushed around the bubble and drawn up from the cluster centre into the wake as it rises (\citealt{Reynolds05,Revaz08,Pope10,Zhang18,Zhang22}).  Gas can be accelerated in the centre of the wake by the pressure gradient and subsequently flows across the underside of the bubble in large eddies, with speeds reaching up to twice the bubble's rise speed.  These simulations typically find bubble terminal rise speeds of $20-50\%$ of the sound speed (\citealt{Churazov00,Zhang18}).  In the wake region of the Ophiuchus cluster, the ICM temperature is $9.4\pm0.3\keV$ and the sound speed is $1570\pm20\kmps$.  The expected terminal rise speed for the bubble is then $\sim300-800\kmps$.

The bulk velocity of the ICM in the bubble's wake is measured here as $-120\pm20\kmps$.  This is measured relative to the mean redshift of 142 cluster members with projected radii to $2\Mpc$.  The distribution of spectroscopic redshifts are statistically indistinguishable from a Gaussian and consistent with a relaxed structure (\citealt{Durret15}).  The mean redshift of the cluster galaxies therefore likely denotes rest within the gravitational potential, and does not appear skewed by merger substructures or the AGN outburst.  The central galaxy's redshift is consistent with rest within the cluster.  In the cluster core, the bulk velocity of the ICM is consistent with rest within the central galaxy and only modestly offset from the surrounding cluster galaxies (\citealt{Fujita25}).  The sharp increase in bulk velocity beneath the large bubble has likely been induced by its creation or subsequent motion.  

Although the bulk velocity appears low compared to simulated bubbles, this likely indicates that the bubble propagates at an angle to the line of sight.  For a true rise speed of $300-800\kmps$, the inclination of the bubble's trajectory with respect to the line of sight is then $65-80^{\circ}$.  For a randomly oriented vector of magnitude $v$, the probability that the component along the line of sight is $v_{\mathrm{los}}$ or smaller is $|v_{\mathrm{los}}|/v$.  For $v=300-800\kmps$ and $v_{\mathrm{los}}=-120\pm20\kmps$, the probability is $15-40\%$ and entirely reasonable.  The relatively high velocity dispersion, compared to the bulk velocity, is consistent with a high terminal velocity compared to the line of sight component.

A high terminal velocity at the upper end of this range is indicated by the radio source age and the sharp X-ray surface brightness edge beneath the bubble.  If the terminal velocity is $300\kmps$, the time taken for the bubble to rise from the cluster centre to a radius of $300\kpc$ (the bubble's centre) is $\sim1\Gyr$.  This is inconsistent with the radio source age of $\sim170\Myr$ inferred from a break in the synchrotron radio spectrum for the relativistic plasma within $200\kpc$ of the X-ray edge (\citealt{Giacintucci25}).  This estimate necessarily includes assumptions, such as a constant and uniform magnetic field and negligible dynamical effects, that may not be applicable across the large spatial scales spanned by the radio bubble in Ophiuchus (e.g. \citealt{Rudnick02}).  However, the sharp X-ray cavity rim (Fig. \ref{fig:chandraimg}) is also inconsistent with a rise time of $\sim1\Gyr$.

From \textit{Chandra} images, \citet{Werner16} estimate the radius of curvature of the cavity to be $r_{\mathrm{cav}}=180\kpc$.  The high sound speed of the relativistic plasma inside the radio lobe should ensure that the pressure within it is roughly uniform.  Across the underside of the bubble, the component of gravity parallel to the surface is then not balanced by a pressure gradient.  In the absence of other forces, ICM blobs in the bubble's rim will move along the interface, towards the radial axis of the bubble.  The initial acceleration of the blobs will be $g\mathrm{sin}\theta$, where $g$ is the acceleration due to gravity and $\theta$ is the angular offset of the blob from the radial axis of the bubble.  The subsequent motion of the blob towards the wake centre can be approximated by simple harmonic motion with $r_{\mathrm{cav}} \left(d^{2}\theta/dt^{2}\right) = - g\theta$.  Ignoring interactions between different infalling blobs, the time it would take for a blob to fall around the rim to the wake centre is roughly a quarter period, where $t_{\mathrm{rim}}\sim0.5\pi\sqrt{r_{\mathrm{cav}}/g}$.  This can equivalently be expressed as $t_{\mathrm{rim}}\sim0.5\pi\sqrt{r_{\mathrm{cav}}R/v_{\mathrm{K}}}$, where $R$ is the distance from the bubble's rim to the cluster centre and $v_{\mathrm{K}}$ is the Kepler speed (at this radius, roughly equivalent to the local sound speed).  For $r_{\mathrm{cav}}=180\kpc$, $R=120\kpc$ and $v_{\mathrm{K}}\sim1400\kmps$, $t_{\mathrm{rim}}\sim140\Myr$ and this is roughly the timescale on which we would expect significant disruption to the rim.  This is clearly inconsistent with a bubble rise time of $\sim1\Gyr$.

In reality, the gas flow beneath the radio bubble is considerably more complicated.  To inflate the radio bubble, the radio plasma injected through the jet must have a greater pressure than the ICM that it displaces.  Initially, the expanding lobe would have probably driven a shock into the ICM and the pressure may have been sufficient to push the ICM beneath the lobe to smaller radii in a backflow.  We note that this may induce sloshing of the ICM in the core, which is indicated by both the X-ray cold fronts (\citealt{Werner16}) and the physical and dynamical offset between the central galaxy and the optical line-emitting gas (\citealt{Edwards09,Hamer12}).  As the lobe expands to larger radii, it encounters lower ICM pressure and the lobe pressure will likely decrease with time (assuming roughly constant jet power).  When the lobe pressure drops below that of the ICM beneath the bubble, the ICM there is accelerated outward by the pressure of the ICM under and around it.  This gives rise to the outflow, or splash, seen in a bubble's wake.  Although the gas flow in the splash is complicated, gravity will cause the ICM to fill in the observed indent on a timescale of roughly $t_{\mathrm{rim}}$.  In summary, the observed sharp bubble rim cannot be sustained for $\sim\Gyr$ timescales.  A higher terminal velocity of $800\kmps$ (or $50\%$ of the sound speed) would ensure a more reasonable rise time of $\sim350\Myr$.

We note that this argument would fail if the cavity's radius of curvature is strongly underestimated.  This would occur if the bubble was highly flattened and observed at an unfavourable viewing angle (with limited extent along the line of sight).  However, from the \textit{Chandra} observations, the X-ray surface brightness decrement in the cavity is $\sim30\%$, which is consistent with a spherical cavity.  We therefore conclude that the bubble is rising at an angle of $\sim80\deg$ to the line of sight and the bubble's terminal velocity is $\sim800\kmps$.  The relatively high velocity dispersion, compared to the bulk motion, is consistent with this picture and suggests that turbulent motions may be limited.


\subsection{Detection of the splash beneath the radio bubble}

The bulk velocity in the cool core is consistent with the central galaxy's velocity and only modestly offset from the mean velocity of the cluster members.  The clear velocity shift and increase in dispersion in the wake are consistent with the motion of the vast radio bubble rising at an angle of $\sim80\deg$.  As discussed in section \ref{sec:terminal}, gas motions beneath the bubble should be dominated by the `splash', where ICM displaced by the bubble is accelerated outward in the wake.  Simulations of these gas motions beneath buoyant radio bubbles show that the induced bulk velocity will be greatest along the radial axis of the bubble (\citealt{Revaz08,Pope10,Zhang18,Zhang22}).  Our analysis of sub-regions across the underside of the radio bubble are consistent with this picture.  The velocity shift in the wake, and the increase in dispersion, are primarily observed in the wake centre region.  A velocity shift is also observed in the SW region, which coincides with a region of enhanced radio emission (Fig. \ref{fig:chandraimg} right) and X-ray surface brightness (\citealt{Werner16}).  We have presumed that the bubble has risen along an axis that runs from the central galaxy to the midpoint of the bright rim, through the wake C region.  However, the velocity shift to the SW suggests that the bubble's expansion and the gas displacement may not have been axisymmetric.

The cool core does not appear strongly disrupted by the inflation of the radio bubble.  \textit{Chandra} observations reveal steep temperature and metallicity gradients that survived this powerful outburst (\citealt{Werner16}).  This suggests that a collimated jet pierced through the core to inflate the bubble at a larger radius.  The observed cold fronts are consistent with gas sloshing along the jet axis, which may have been induced by the splash in the wake of such a large bubble.  This scenario would explain why bulk motion and turbulence in the core remain remarkably low, and the velocity shift occurs at larger radii beneath the bubble.  Initially, the ICM was at rest in the gravitational potential, given by the mean velocity of the cluster members.  A powerful jetted outburst inflated the bubble beyond the core.  The bubble's expansion and subsequent buoyant motion induced the observed bulk velocity in the wake and the modest sloshing of the cool core.  If the bubble was inflated beyond the core, the uplifted material in the wake will have comparable gas properties to the ambient ICM and we would not expect to detect bright, radial, soft X-ray filaments or a clear extension of low entropy, metal-rich material along the bubble axis.  However, we note that entropy maps from the \textit{Chandra} observations hint at lower entropy material, relative to the ambient, in the region of enhanced X-ray surface brightness along the SW side of the radio bubble (Fig. \ref{fig:chandraimg} right, \citealt{Werner16}).


Although cold fronts may instead have been induced by a merger, we now consider this unlikely.  \citet{Werner16} considered a merger origin for the cavity rim because similar concave X-ray surface brightness discontinuities, known as bays, appear at certain stages of subcluster infalls (e.g. \citealt{Walker17}).  However, the radio emission associated with these features is then constrained behind the concave edge.  In Ophiuchus, the cavity beyond the edge is instead filled with radio emission, which indicates a radio bubble (\citealt{Giacintucci20}).  Furthermore, the X-ray surface brightness decrement within the cavity of $\sim0.3$ is consistent with a spherical cavity devoid of ICM.


\subsection{Heating via turbulent dissipation}

With an associated jet energy of $5\times10^{61}\erg$ (\citealt{Werner16}), the radio bubble in the Ophiuchus cluster is one of the most energetic known (\citealt{Giacintucci20}).  For our new estimate of the rise speed of $800\kmps$ ($50\%$ of the sound speed), the jet power is then $\sim5\times10^{45}\ergps$.  The jet energy must be dissipated within the cooling radius to suppress radiative cooling of the ICM.  For the Ophiuchus cluster, the cooling radius, defined as the radius at which the radiative cooling time of the gas drops below $7.7\Gyr$ (time since the last major heating event $z\sim1$, eg. \citealt{Rafferty06}), is $\sim50\kpc$ ($85\arcsec$).  This does not extend far beyond the bright core and is essentially covered by the circular sky region in Fig. \ref{fig:skyreg}, which lies beneath the wake.  The total X-ray luminosity ($0.01-50\keV$) within the cooling radius is the cooling luminosity $L_{\mathrm{cool}}=8.17\pm0.08\times10^{43}\ergps$ and the thermal energy radiated in $7.7\Gyr$ is $1.98\pm0.02\times10^{61}\erg$.  Over the $350\Myr$ rise time of the bubble, the energy radiated within the cooling radius is still substantial at $9.02\pm0.09\times10^{59}\erg$. The jet power is a factor of order $50$ times the cooling luminosity, suggesting that the jets are hugely overpowered, but there is no evidence of strong heating in the ICM.  \citet{Werner16} compare radial entropy profiles in five sectors and show that these are essentially consistent within the errors, with no strong variations in the cavity or neighbouring sectors.

To prevent overcooling in the core, the initial bubble expansion and propagation must have generated sufficient turbulent-dissipation heating in this region or, later on, energy generated in the outer wake propagates into the core.  By assuming that gas sloshing and velocity gradients contribute minimally to the velocity dispersion, we can calculate upper limits on the turbulent kinetic energy.  By integrating the deprojected electron density profile determined from \textit{Chandra} observations, we can calculate the mass of hot gas in the core (to a radius of $55\kpc$) and in the wake region of the XRISM pointing ($\sim55-130\kpc$ in radius and $200-275\deg$, where angles are measured northward from west).  We assume that the extent of the wake along the line of sight is the same as the extent on the sky.  The correction for inclination is insignificant for an estimated bubble trajectory only $10\deg$ from the plane of the sky.  The hot gas mass in the core is $3.1\pm0.3\times10^{11}\Msun$ and in the wake is $\sim0.9\pm0.1\times10^{11}\Msun$.  Assuming isotropic turbulence, with $\sigma=135\pm10\kmps$ in the core and $\sigma=210\pm20\kmps$ in the wake, the turbulent kinetic energy, $\frac{3}{2}M_{\mathrm{gas}}\sigma^2$, is up to $1.7\pm0.2\times10^{59}\erg$ in the core and $1.2\pm0.2\times10^{59}\erg$ in the wake.  In the core, the turbulent kinetic energy is only $1\%$ of the thermal energy radiated within the cooling radius in the cooling time of $7\Gyr$.  This is comparable to the low fraction of $2.5\%$ found in Hydra A (\citealt{Rose25}) and consistent with the low ratios of kinetic to thermal energy found in Abell 2029, Centaurus and Perseus (\citealt{XRISM25A2029, XRISM25Centaurus, XRISM25Perseus}).  The turbulent kinetic energy even falls short, by a factor of 5, compared to the thermal energy radiated over the rise time of the bubble.

Overcooling in the ICM can still be prevented if the turbulent energy is rapidly thermalized and replenished.  To balance radiative cooling in Ophiuchus, this must occur on a timescale of $70\Myr$.  With only a single radio bubble visible, we estimate a lower limit on the AGN duty cycle given by the bubble's rise time of $350\Myr$.  This appears in clear tension with the required replenishment timescale.  Turbulent energy generated in the wake region may propagate into the core to provide additional heating.  However, even for an optimistic assumption on the large-scale diffusion rate (eg. \citealt{Vazza12,Li25}, McNamara et al. in prep.), it would take longer than the age of the cluster for this turbulent energy to travel $100-150\kpc$ across the wake and core.  

The XRISM pointing was targeted at the wake region, where the strongest bulk motion is expected, and may have missed turbulent gas motions induced by the bubble over a wider region.  From the bubble's volume and the electron density profile, we estimate that the bubble has displaced $10^{12}\Msun$ of hot gas (roughly equivalent to the gas mass within a complete spherical shell that spans the radial range of the wake region from $55$ to $130\kpc$).  The turbulent kinetic energy is then $10^{60}\erg$.  This assumes that a similar level of turbulent motion has been induced in this larger volume and is therefore certainly an overestimate.  The velocity dispersion appears to peak in the wake centre and then decrease beyond this.  The short replenishment timescale remains a key problem, particularly given the much larger volume.


Following \citet{Rose25}, we can estimate the dissipative heating rate from the turbulent kinetic energy as 

\begin{equation}
    \dot{E}_{\mathrm{heat}} \sim \frac{3}{2} \frac{\sigma_{\mathrm{l}}^3}{l} M_{\mathrm{gas}}
\end{equation}


\noindent where $M_{\mathrm{gas}}$ is the hot gas mass and $\sigma_{\mathrm{l}}$ is the turbulent velocity measured on a corresponding length scale $l$.  We assume a single Kolmogorov-like turbulent cascade down from the outer scale to the dissipation scale (see e.g. \citealt{Werner16}).  

The ICM is stable against convection; a perturbed ICM blob will oscillate rather than continuing to rise or sink.  This generally limits the outer scale in the radial direction to $\omega_{\mathrm{BV}}l\lesssim \sigma$, where $\omega_{\mathrm{BV}}$ is the Brunt-V\"ais\"al\"a frequency.  Whilst turbulence is almost certainly anisotropic, and the outer scale may be larger in the transverse direction, the shortest scale should dictate the dissipation rate.  The Brunt-V\"ais\"al\"a frequency (\citealt{Cox80}) is given by

\begin{equation}
\omega_{\mathrm{BV}} = \sqrt{\frac{1}{\gamma}\frac{d\mathrm{ln}K(r)}{d\mathrm{ln}r}\frac{g}{r}}
\end{equation}

\noindent where $\gamma=5/3$ for the ICM and $K=k_{\mathrm{B}}Tn^{-2/3}$ is the entropy index.  Substituting for the gravitational acceleration $g=v_{\mathrm{K}}^2/R$, the outer scale is given by

\begin{equation}
l\sim\frac{\sigma_{\mathrm{l}}R}{v_{\mathrm{K}}}\sqrt{\frac{\gamma}{\frac{d\mathrm{ln}K(r)}{d\mathrm{ln}r}}}
\end{equation}

\noindent and the dissipation rate is then

\begin{equation}
    \dot{E}_{\mathrm{heat}} \sim \frac{3}{2} \frac{v_{\mathrm{K}}}{R}\sigma_{\mathrm{l}}^2M_{\mathrm{gas}}\sqrt{\frac{1}{\gamma}\frac{d\mathrm{ln}K(r)}{d\mathrm{ln}r}}.
\end{equation}

\noindent Additional factors are expected to be of order unity.  From the \textit{Chandra} observations, the logarithmic derivative of the entropy index, $d\mathrm{ln}K(r)/d\mathrm{ln}r$ is $0.90\pm0.07$ and, assuming hydrostatic equilibrium, $v_{\mathrm{K}}\sim500\kmps$.  The outer scale is then $\sim20\kpc$.  The corresponding dissipative heating rate in the core is then $\sim3\times10^{43}\ergps$.  For gas dynamics within the core that are strongly influenced by earlier expansion of the radio bubble, the scale associated with the measured turbulent velocity could smaller (as low as $10\kpc$, e.g. \citealt{Shin16}) and the dissipative heating rate could correspondingly be up to a factor of two higher.  However, the short replenishment timescale requires motions induced by early bubble expansion to have dissipated and a larger scale is then more appropriate. 


The heating rate in the core falls short of the cooling luminosity by a factor of 3.  Furthermore, we have assumed a negligible contribution to the velocity dispersion from unresolved velocity gradients.  Smooth velocity gradients spanning a few hundred km/s are frequently observed in the cool gas filaments lifted in the wakes of radio bubbles (\citealt{Hatch07,Russell19, Olivares19}).  Similar velocity gradients measured for both the cool filaments and ICM in the Hitomi observation of the Perseus cluster suggest the cool and hot gas phases co-move (\citealt{Hitomi16}).  The observed variation in bulk motion across the wake in Ophiuchus suggests there could be unresolved gas flows in this region, as expected from simulations.

In summary, despite the vast power of the radio bubble in the Ophiuchus cluster, turbulent-dissipation heating would have difficulty preventing overcooling in the cluster core.  At $2.4\times10^{60}\erg$, the energy of the inner bubbles in the core of Hydra A is roughly an order of magnitude below that of the bubble in Ophiuchus and yet the turbulent kinetic energy in the core is at least a factor of 10 higher (\citealt{Rose25}).  Despite the range in jet power spanned by early XRISM targets, from M87 to the Ophiuchus cluster, modest turbulent velocities are prevalent and, as yet, show no evidence of the expected correlation with jet power.  This could straightforwardly be explained by attributing a significant fraction of the velocity dispersion to unresolved velocity gradients rather than turbulence.  However, turbulent dissipation would then struggle to prevent overcooling in the Ophiuchus cluster core.



\section{Conclusions}

Using a $100\ks$ Resolve observation of the Ophiuchus cluster, we measured the bulk motion and velocity dispersion in the wake of the largest radio bubble on the sky.  We detect a clear velocity shift and increase in velocity dispersion from the cluster centre to the wake region beneath the radio bubble.  The gas velocities are low with bulk motion in the wake of only $-120\pm20\kmps$.  We demonstrate that the bubble's trajectory is likely inclined with respect to the line of sight.  Simulations indicate terminal velocities of $20-50\%$ of the sound speed, which corresponds to $300-800\kmps$, and bubble rise times of $350-1000\Myr$.  The stability of the X-ray surface brightness edge denoting the cavity rim and the age of the radio emission within the bubble imply a short rise time in this range.  The bubble is therefore rising at an angle of $\sim80\deg$ to the line of sight with a terminal velocity of $\sim800\kmps$.  The relatively high velocity dispersion, compared to the bulk motion, is consistent with this picture and suggests that turbulent motions in the wake may be limited.

Gas motions beneath the bubble should be dominated by the splash, where ICM displaced by the bubble is accelerated outward in the wake.  The clear shift and increase in dispersion in the wake are consistent with this scenario.  Furthermore, our analysis of sub-regions across the underside of the radio bubble find that the velocity shift and increase in dispersion are primarily detected in the centre of the wake.  Simulations of gas flow beneath buoyant radio bubbles show that the induced bulk velocity will be greatest along the bubble's radial axis, as observed.

Steep temperature and metallicity gradients in the cool core suggest that the jet pierced through this region without disrupting it and inflated the vast bubble at a larger radius.  This explains why bulk motion and turbulence in the core remain low, and the velocity shift and dispersion increase occur at larger radii beneath the bubble.  The modest shift in bulk motion in the cool core, relative to the cluster member galaxies, may be due to gas sloshing induced by the expansion of the vast bubble beyond the core.  The expanding lobe would have likely driven a shock into the ICM and the pressure may have been sufficient to push the ICM beneath the bubble to smaller radii.  The observed cold fronts, and the physical and dynamical offset between the central galaxy and its cool gas nebula, are consistent with this scenario.

Despite hosting of the most energetic outbursts known, the velocity dispersions in the core and wake of the Ophiuchus cluster are only $135\pm10\kmps$ and $210\pm20\kmps$, respectively.  The turbulent kinetic energy in the cluster's core region is only $1\%$ of the thermal energy radiated on the cooling timescale and only $20\%$ of the energy radiated over the bubble's rise time.  Although the bubble may generate turbulent energy throughout the large wake region, this propagates too slowly to provide significant additional heating on the timescale required by the rapidly cooling core.  We estimate the turbulent-dissipation heating rate to be a factor of $\sim3$ below the cooling luminosity.  If unresolved velocity gradients contribute to the measured velocity dispersion, the heating rate will be even lower.  We therefore suggest that turbulent-dissipation heating would struggle to prevent rapid gas cooling in the cluster core.

\section*{Acknowledgements}

We thank the dedicated team of scientists and engineers who worked on the XRISM mission.  We are grateful to Simona Giacintucci for supplying the $210\MHz$ GMRT radio contours shown in Fig. \ref{fig:chandraimg} (right).  We thank the reviewer for their helpful and constructive comments.  HRR acknowledges support from an Anne McLaren Fellowship from the University of Nottingham and funding from a Leverhulme Trust Research Leadership Award.  NW was supported by the GACR grant 21-13491X.  This research has made use of data from the \textit{Chandra} X-ray Observatory and software provided by the \textit{Chandra} X-ray Center (CXC).  The \textit{Chandra} data sets that we use are contained in the \textit{Chandra} data collection at: https://doi.org/10.25574/cdc.528.  This work used observations obtained with XMM–Newton, an ESA science mission funded by ESA Member States and the USA (NASA).  Many of the plots in this paper were made using the Veusz software, written by Jeremy Sanders.  This research made use of \textsc{astropy}, a community-developed core Python package for Astronomy (\citealt{Astropy22}).

\section*{Data Availability}
The XRISM data described in this work will be available in the NASA HEASARC (https://heasarc.gsfc.nasa.gov/docs/xrism/archive/).  Processed data products detailed in this paper will be made available on reasonable request to the author.



\bibliographystyle{mnras}
\bibliography{refs} 








\bsp	
\label{lastpage}
\end{document}